\documentclass[12pt]{article} 
\usepackage{graphicx}
\usepackage{amsmath,amssymb}
\usepackage{longtable}

\begin{document}
\baselineskip 21pt

\bigskip

\centerline{\Large \bf Structure Decomposition}
\centerline{\Large \bf for the Luminous Disk Galaxies in the NGC 524 Group}

\bigskip

\centerline{\large M.A. Ilyina$^1$ and O.K. Sil'chenko$^1$}

\noindent
{\it Sternberg Astronomical Institute of the Lomonosov Moscow State University, Moscow, Russia}$^1$

\vspace{2mm}
\sloppypar 
\vspace{2mm}

\bigskip

{\small 
\noindent
Members of the NGC 524 group of galaxies are studied using data obtained on the 6m 
telescope of the Special Astrophysical Observatory of the Russian Academy of Sciences, 
with the SCORPIO reducer in an imaging mode. Surface photometry has been carried out 
and parameters of the large-scale galactic components -- disks and bulges -- have been 
determined for the six largest galaxies of the group. A lower than expected percentage of bars 
and high percentage of ring structures were found. Integrated $B-V$ colours for a 
hundred of dwarf galaxies in the vicinity (within 30 kpc) of the six largest galaxies of 
the group have been measured. A considerable number of blue irregular galaxies with 
ongoing star formation is found among dwarf satellites of the lenticular galaxies 
of the group. The luminosity function for dwarf galaxies of the group suggests that 
the total mass of the group is not very high, and that the X-ray emitting gas observed 
around NGC 524 relates to the central galaxy and not to the group as a whole. 
}

\clearpage

\section{INTRODUCTION}

Groups of galaxies are very suitable places to study there the role of
interactions concerning the evolution of galaxies due to external factors. 
The probability of the development of gravitational tides during close
passages is the highest for the galaxies in groups, due to both the presence 
of close neighbours and the low relative velocities of intragroup movements. 
Groups, especially massive ones, often contain hot (X-ray), diffuse
gas, which interacts with the cool gas of the galaxies themselves; such 
gas-dynamical interaction is usually considered as a reason for the dearth 
of cool gas in disk galaxies that are members of groups.

So-called ``external secular evolution'' (see the review by \cite{kk04}) 
is a slow change of the global properties of galaxies caused by external 
factors. Secular evolution leaves certain imprints on the large-scale 
structures of galaxies: the radial density profiles of their disks become 
more complex, and additional structural components, such as bars, rings,
and circumnuclear disks, arise in central regions. The dynamical properties 
of the stellar subsystems also change: the stellar disk becomes thicker (hotter)
during the evolution, the vertical velocity dispersion of its stars increases, etc. 
Therefore, the origin of the large-scale structures of galaxies can be studied 
most effectively using a complex approach, beginning with the surface
photometry, and then more clearly identifying the nature of the structural components 
with the spectral observations and kinematic measurements. In the present
paper, which is the first in the series devoted to the nearby, rich NGC 524 group, 
we are starting with the surface photometry for a number of luminous member galaxies.
The photometric data were obtained at the 6-m telescope of the Special Astrophysical Observatory
(SAO) of the Russian Academy of Sciences. 

The NGC 524 group is well known, and has been studied for a long time, 
due to its location in a relatively sparsely populated area in the Northern
sky. It was first cataloged by Geller and Huchra \cite{gellerhuchra}, when only 
eight galaxies were identified as members. Later, Vennik \cite{vennik} found 
18 luminous members and 13 dwarfs by undertaking an analysis of Palomar Atlas
images using the force hierarchy clustering method, and Brough et al. \cite{brough} 
identified 16 members using the ``friends-of-friends'' algorithm. The NGC 524
group lists also 16 bright members in the last catalog of nearby galaxy groups by 
Makarov and Karachentsev \cite{makkar}. Most of the bright members of the group 
are classified as lenticular galaxies; the fraction of early-type galaxies is 
estimated to be $0.56 \pm 0.15$ over the entire luminosity range \cite{brough}. 
The velocity dispersion of the group galaxies  is about 180--190 km/s \cite{vennik,brough},
making the group fairly massive ($1.7 \cdot 10^{13}\, M_{\odot}$) and compact 
($r_{500} = 0.26$~Mpc \cite{brough}, $r_{500} = 0.42$~Mpc \cite{gems}).
The ROSAT mission detected the X-ray emitting gas in the NGC 524 group, 
but since the radius of the detected X-ray spot is less than 60 kpc, 
the hot X-ray gas is thought to belong to the central galaxy
of the group, NGC~524, and not globally to the group \cite{xray}.

We have studied the structure of the six brightest galaxies of the group -- four lenticulars 
and two early-type spirals -- in detail. Their global characteristics retrieved 
from extragalactic databases are given in Table~1.

\begin{table*}
\scriptsize
\caption[ ] {Global characteristics of the galaxies under consideration}
\begin{flushleft}
\begin{tabular}{lcccccc}
\hline\noalign{\smallskip}
Galaxy & NGC~502 & NGC~509 & NGC~516 & NGC~518 
& NGC~524 & NGC~532 \\
Morph.type (NED$^1$) & SA$0^0$(r) & S0? & S0 & Sa: &
SA$0+$(rs) & Sab? \\
$R_{25},\, ^{\prime \prime}$  (RC3$^2$) & 34 & 43 & 39 & 52 
& 85 & 75 \\
$B_T ^0$ (RC3) & 13.57 & 14.20 & 13.97 & 13.56 & 11.17 & 12.91 \\
$M_B$ (RC3$+$NED) & --18.3 & --17.7 & --17.9 & --18.3 & --20.7 & --19.0 \\
$(B-V)$ (RC3) & 0.91 & -- & -- & -- & 1.00 & 0.80 \\
$V_r$, km/s (NED) & 2489 & 2274 & 2451 & 2725 & 2379 & 2361 \\
Group center separation $^{3,2}$, kpc & 282 & 151 & 68 & 101 & 0 & 126 \\
Distance to the group$^4$, Mpc & \multicolumn{6}{c}{24} \\
\hline
\multicolumn{7}{l}{$^1$\rule{0pt}{11pt}\footnotesize
NASA/IPAC Extragalactic Database}\\
\multicolumn{7}{l}{$^2$\rule{0pt}{11pt}\footnotesize
Third Reference Catalogue of Bright Galaxies \cite{rc3}}\\
\multicolumn{7}{l}{$^3$\rule{0pt}{11pt}\footnotesize
Sengupta \& Balasubramanyam (2006), \cite{sengupta}}\\
\multicolumn{7}{l}{$^4$\rule{0pt}{11pt}\footnotesize
Tonry et al. (2001), \cite{tonry}}\\
\end{tabular}
\end{flushleft}
\end{table*}

\section{OBSERVATIONS}

The photometric observations analyzed here were fulfiled at the prime focus of the 6-m telescope 
of the Special Astrophysical Observatory of the RAS with the SCORPIO reducer \cite{scorpio} 
operating in a direct-imaging mode.
The CCD-detector EEV 42-40 had a $2048 \times 2048$ format with a pixel size of 13.5 $\mu$m;
the readout was performed in a double-binned regime, yielding a sampling of 
$0.35^{\prime \prime}- 0.36^{\prime \prime}$ per pixel.  The total field of view of the 
reducer was $6.1^{\prime} \times 6.1^{\prime}$. The observations were carried out 
in the standard Johnson $B$ and $V$ bands. An exposure of twilight sky was used as a flat field.
A detailed log of observations is presented in Table 2. Observations of the NGC 524 group were
performed on the night of 21/22 August, 2007, under photometric conditions with a seeing of 
about $2^{\prime \prime}$.
We took $BV$-band exposures of six fields centered onto the brightest galaxies of the group: NGC 524,
NGC 502, NGC 509, NGC 516, NGC 518, and NGC 532. We used the exposures of the central galaxy of the 
group, NGC 524, as a photometric standard; the HYPERLEDA database contains a good set of aperture
photoelectric data for this galaxy reduced to the standard Johnson system.

\begin{table}
\caption{Photometric observations of the NGC~524 group}
\begin{flushleft}
\begin{tabular}{|l|c|r|c|}
\hline\noalign{\smallskip}
Field center & Band & Total exposure, s & Seeing \\
\hline
NGC 502 & $B$ & 540 & $2.3^{\prime \prime}$ \\
NGC 502 & $V$ & 240 & $2.1^{\prime \prime}$ \\
NGC 509 & $B$ & 540 & $2.0^{\prime \prime}$ \\
NGC 509 & $V$ & 300 & $2.0^{\prime \prime}$ \\
NGC 516 & $B$ & 540 & $2.2^{\prime \prime}$ \\
NGC 516 & $V$ & 240 & $2.0^{\prime \prime}$ \\
NGC 518 & $B$ & 540 & $1.9^{\prime \prime}$ \\
NGC 518 & $V$ & 240 & $2.1^{\prime \prime}$ \\
NGC 524 & $B$ & 60  & $1.8^{\prime \prime}$ \\
NGC 524 & $V$ & 30 & $2.3^{\prime \prime}$ \\
NGC 532 & $B$ & 360 & $1.8^{\prime \prime}$ \\
NGC 532 & $V$ & 180 & $2.3^{\prime \prime}$ \\
\hline
\end{tabular}
\end{flushleft}
\end{table}

\section{STRUCTURE OF THE LARGE DISK GALAXIES IN THE NGC 524 GROUP}

\begin{figure*}
\resizebox{\hsize}{!}{\includegraphics{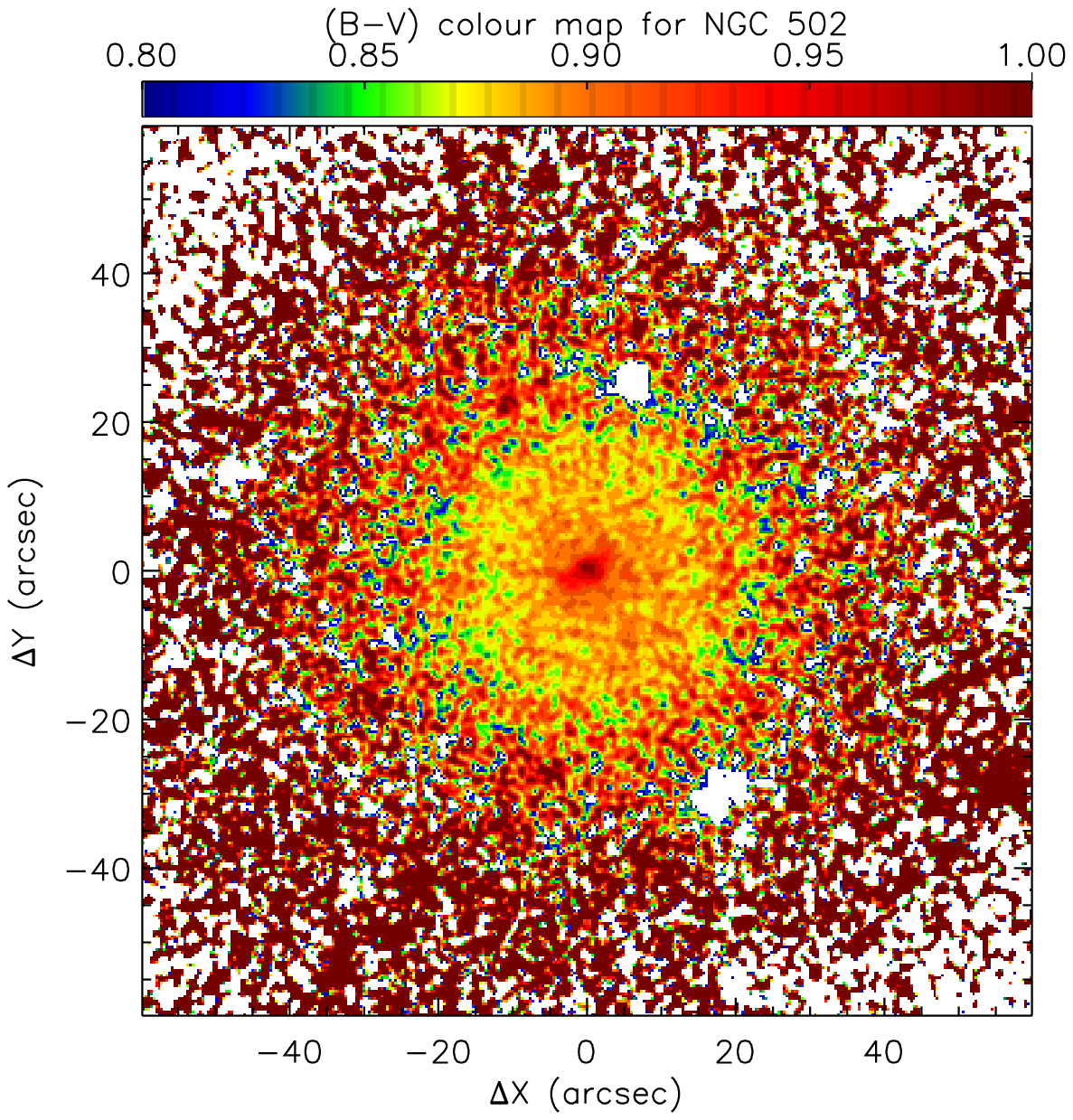}
\includegraphics{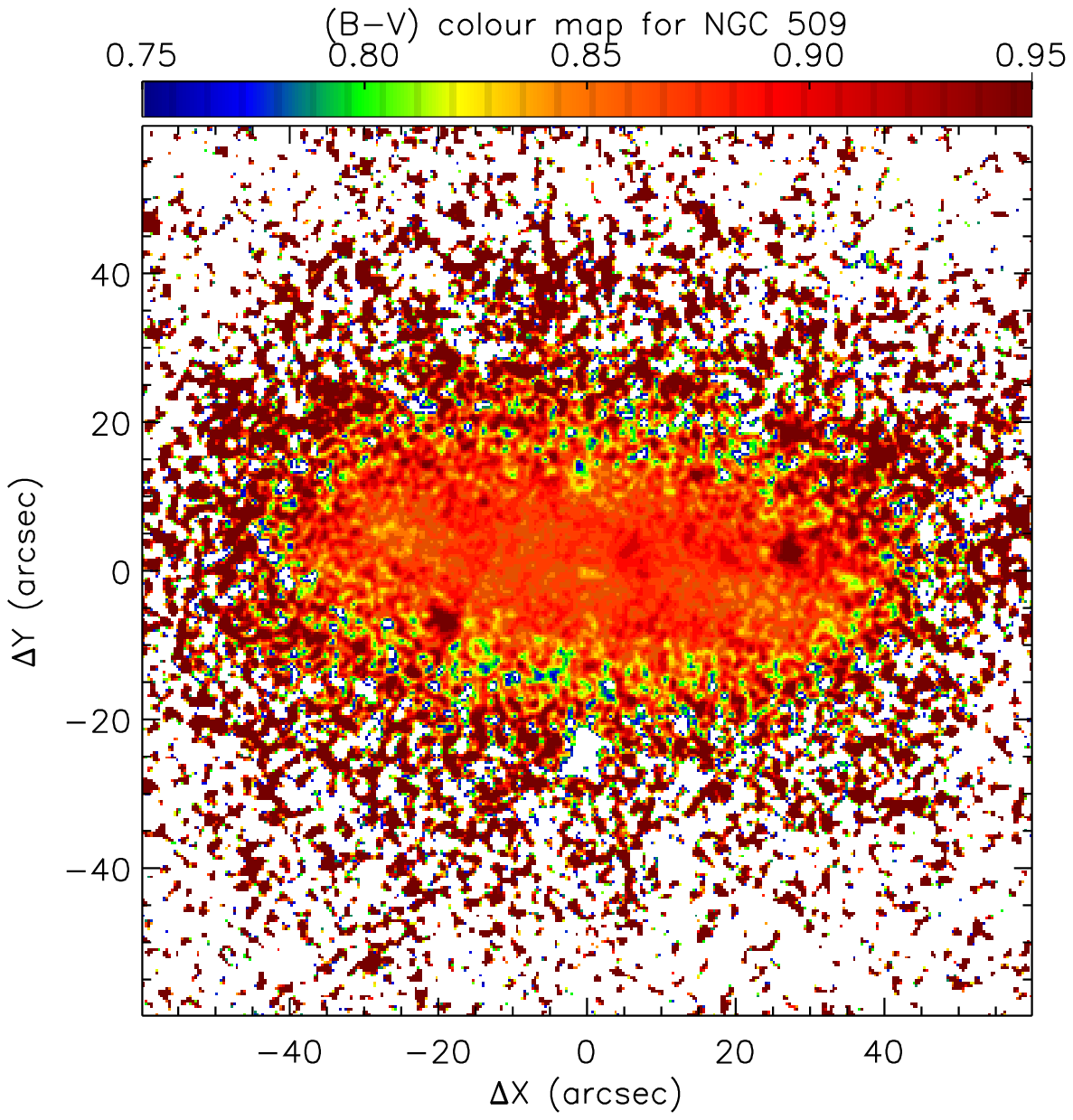}}
\resizebox{\hsize}{!}{\includegraphics{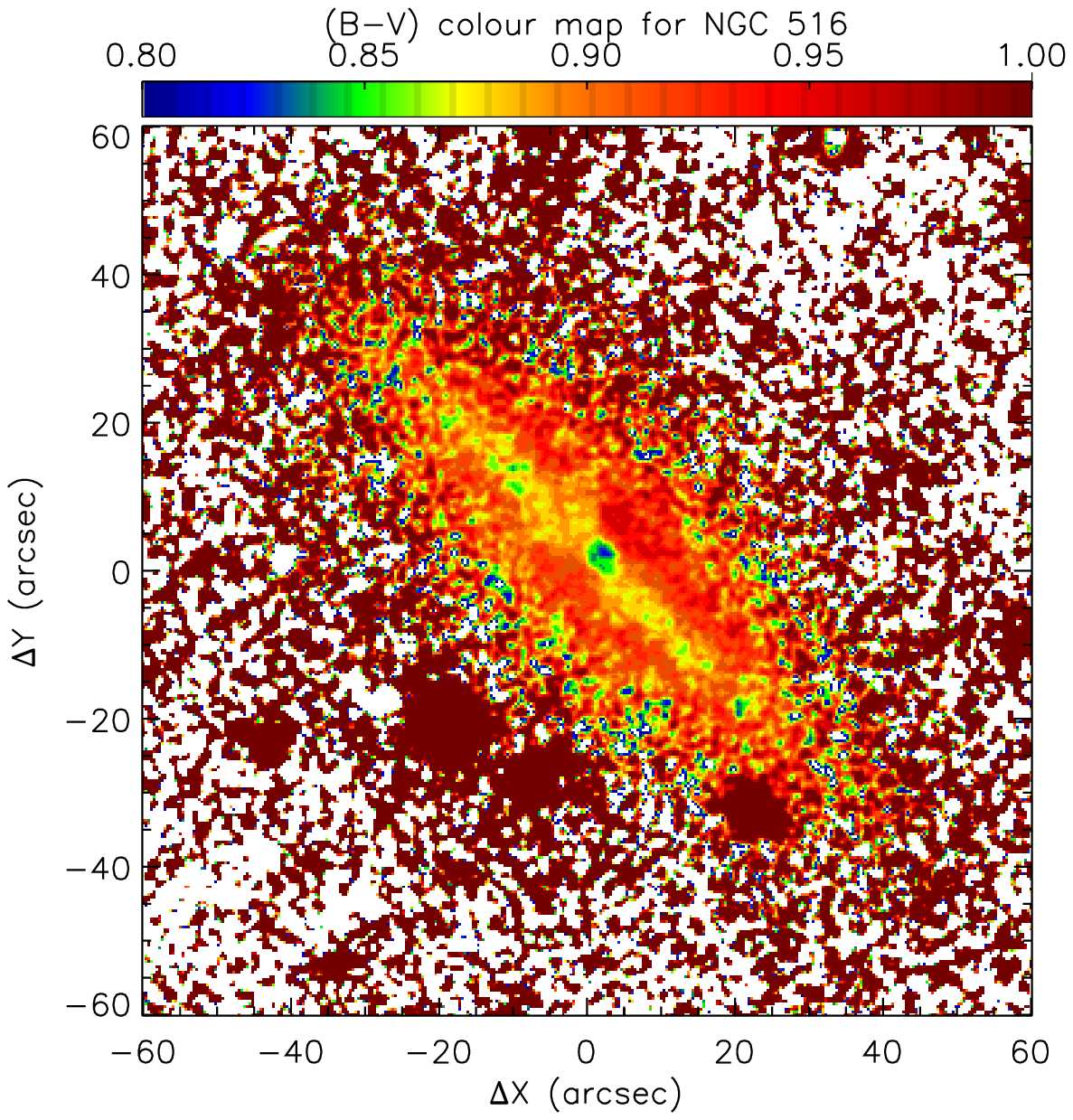}
\includegraphics{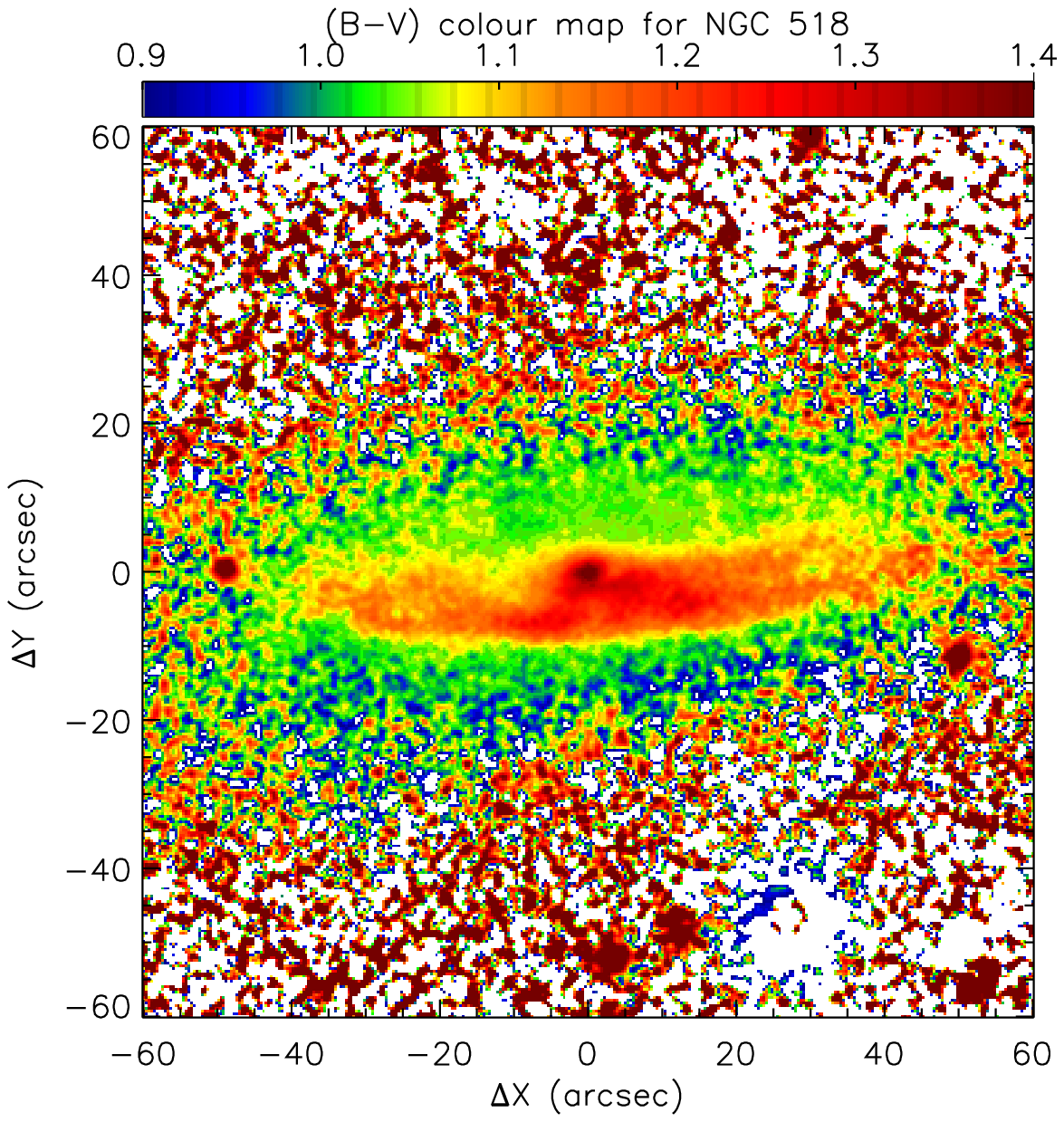}}
\resizebox{\hsize}{!}{\includegraphics{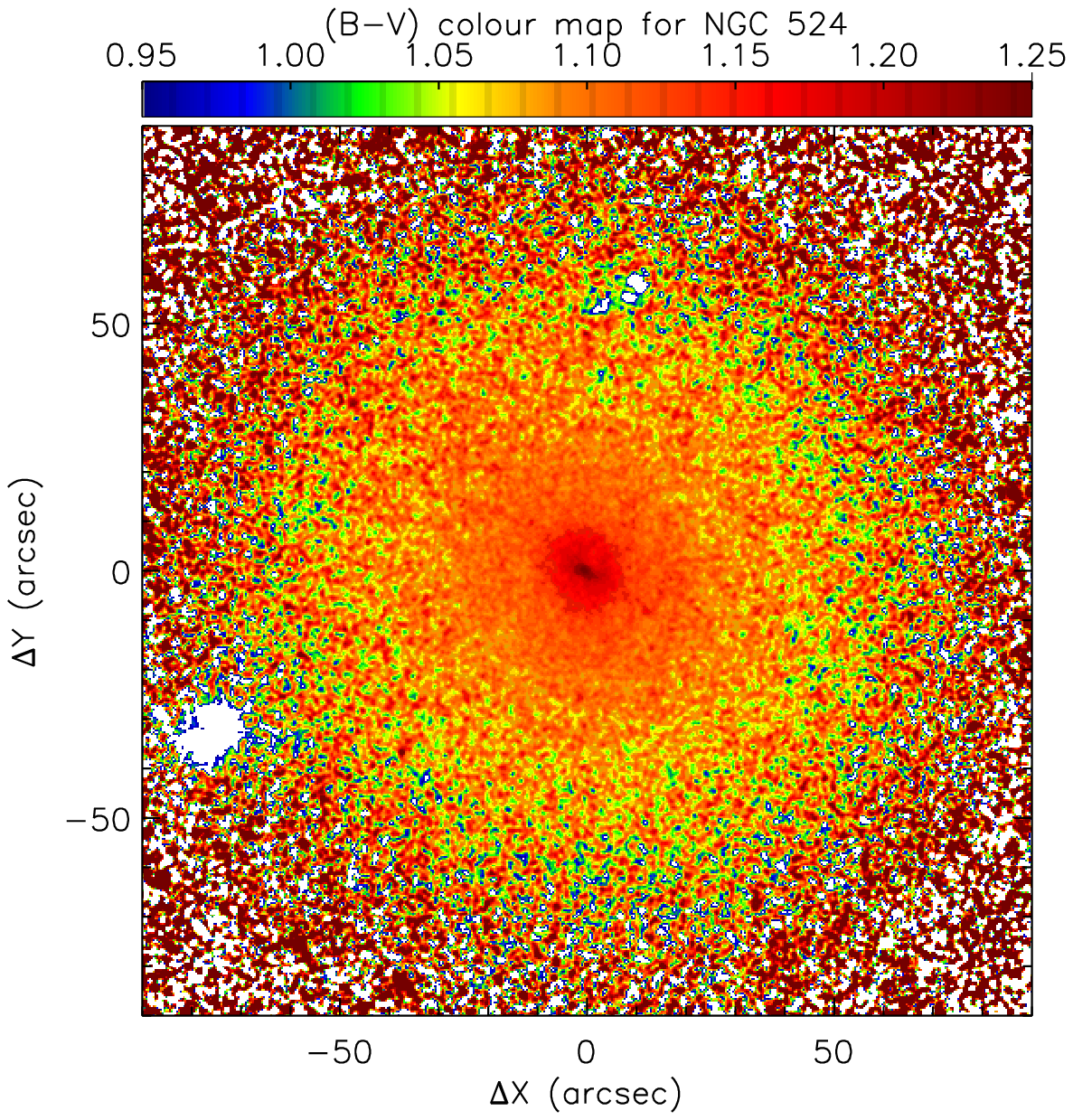}
\includegraphics{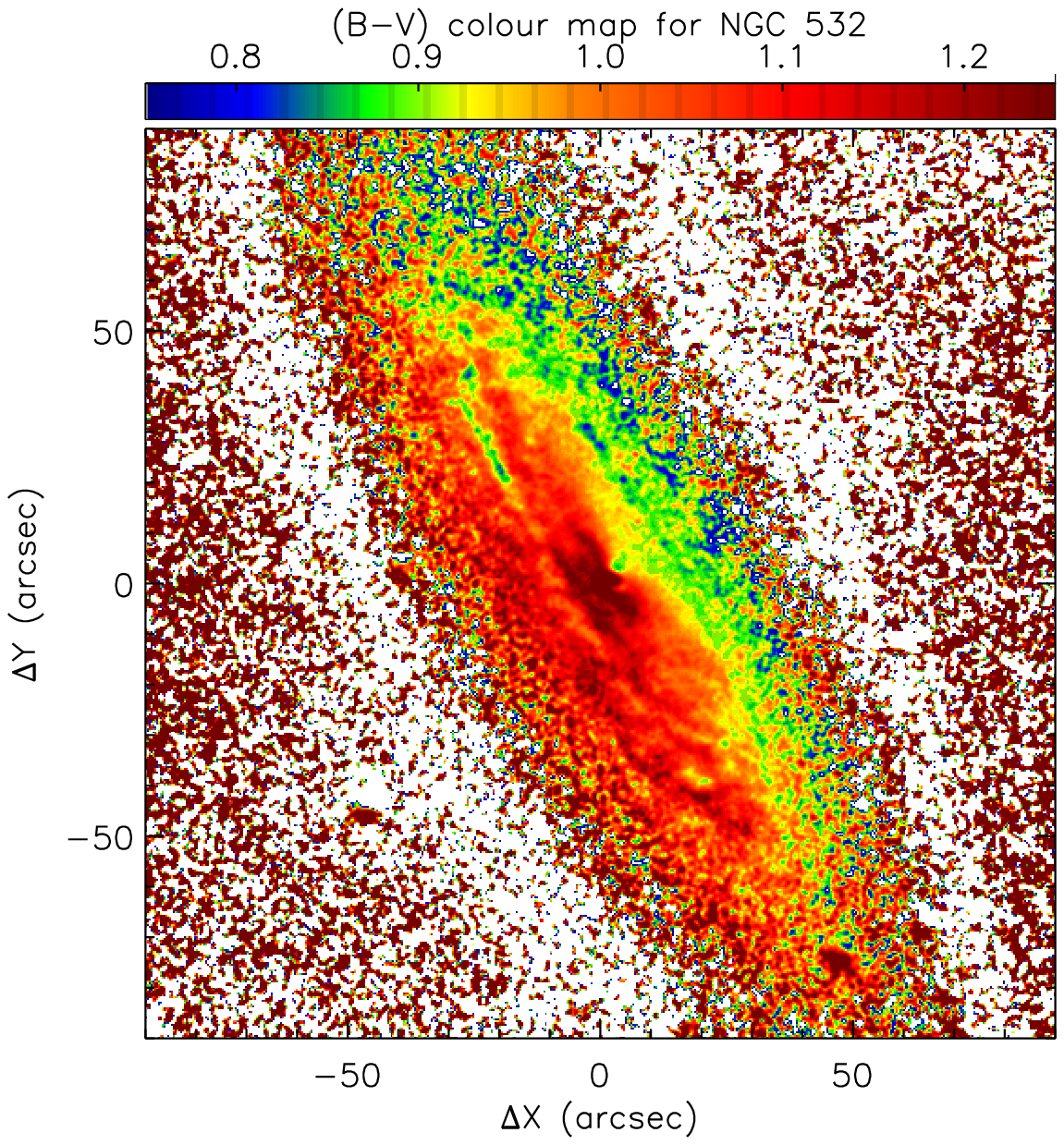}}
\caption{$B-V$ colour maps for the six large galaxies of the NGC 524 group.}
\label{bvmaps}
\end{figure*}

Figure~\ref{bvmaps} presents $B-V$ colour maps of the six brightest disk galaxies of the group. 
The lenticular galaxies NGC 524 and NGC 502 are viewed nearly face-on, and colour maps 
reveal very red nuclei, with the colours becoming bluer outward in a monotonic,
axially symmetric fashion representing the classical radial colour gradient in early-type 
galaxies, usually interpreted as a gradient of the metallicity of the stellar population.
On the contrary, the lenticular galaxies NGC 509 and NGC 516 are viewed edge-on. Unlike
NGC 524 and NGC 502, their nuclei are bluer than the main bodies of the galaxies. Outside 
the nucleus, NGC 509 shows uniformly red colours without any noticeable trend with radius, 
while NGC 516 has a thin bluish disk embedded into a redder structure that can be taken 
as a thick stellar disk. The colour maps for the spiral galaxies NGC 518 and NGC 532 
display strong dust lanes projected onto the bulges, which relate to the spiral arms
of these galaxies; the shift of the projections of the dust arms relative to 
the nuclei along the minor axes confirms our isophotal analysis conclusion,
that these galaxies are not viewed strictly edge-on, but under the inclination of 
approximately $80^{\circ}$.

\begin{figure*}
\resizebox{\hsize}{!}{\includegraphics{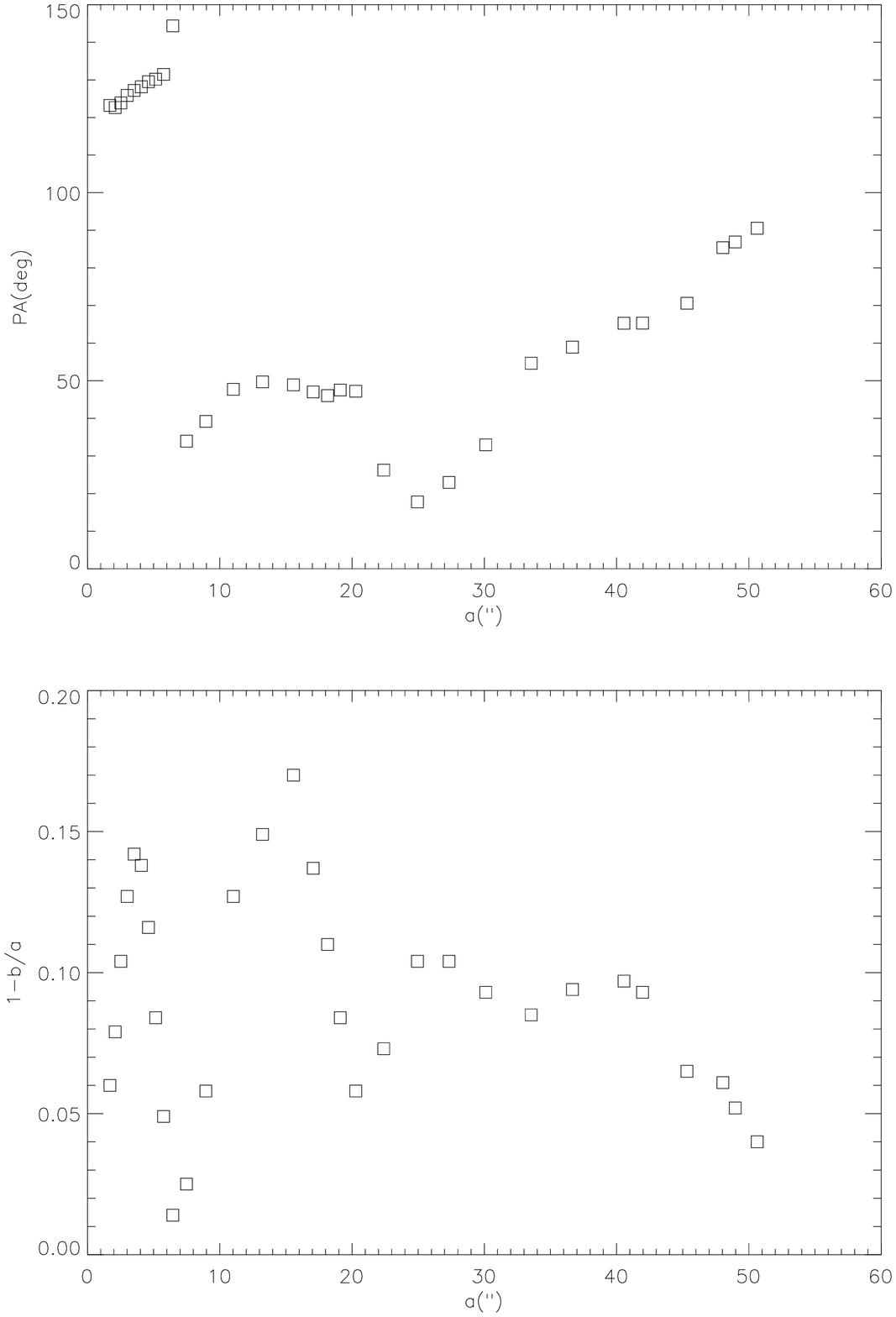}
\includegraphics{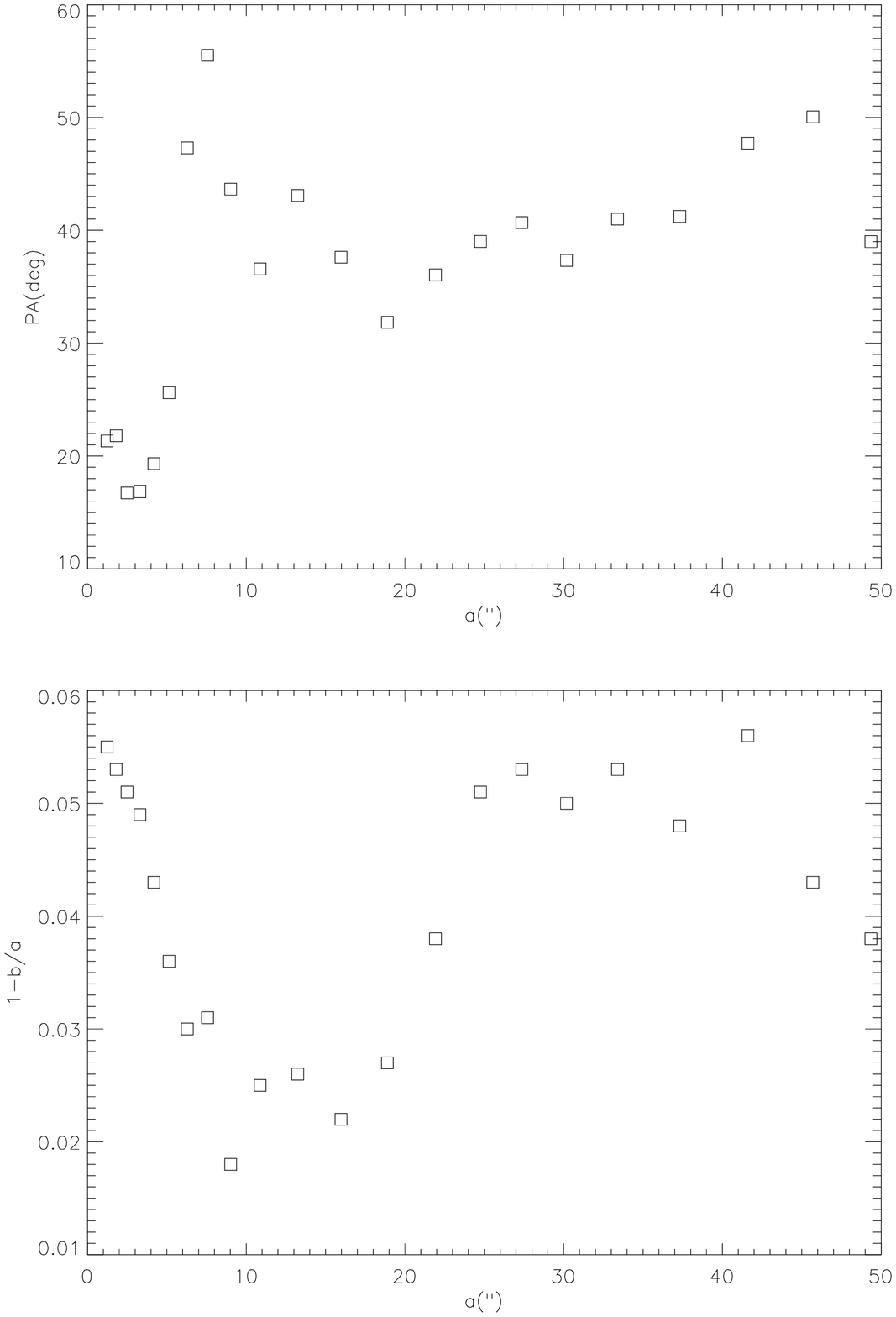}
\includegraphics{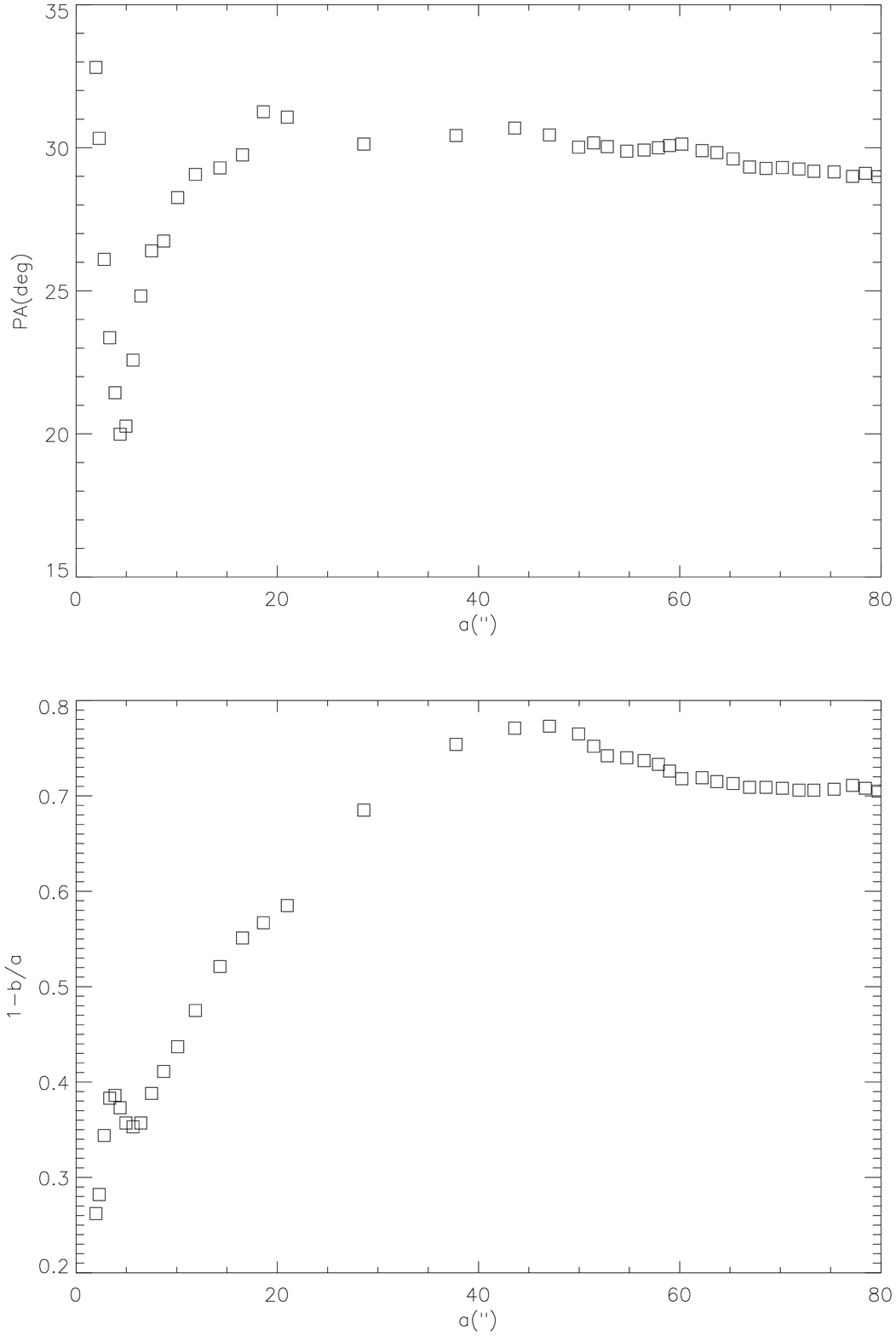}}
\caption{Results of our isophotal analysis of the $V$ images for {\it a} -- NGC 502, {\it b} -- NGC 524, 
and {\it c} -- NGC 532. The radial dependencies of the major-axis position angle (top) and 
the ellipticity of the isophotes (bottom) are shown.}
\label{iso}
\end{figure*}

We performed an isophotal analysis and determined the major-axis position angles and 
ellipticities of the isophotes as a function of radius for all six galaxies in both filters. 
As a rule, the isophote ellipticity reaches a plateau at some distance from the center 
and does not change further with radius. We assume that this level, $e \equiv 1 - b/a$, 
is related to the inclination of the galactic disk to the line of sight $i$ as following

$$
\sqrt{\frac{2e-e^2}{1-q_0^2}}=\sin i
$$

\noindent
where $q_0$ takes into account the ``thickness'' of the disk and is equal to the scale height and
scale length ratio, $z_0/h$.  The radius where the ellipticity reaches a plateau borders
the region where the thin outer stellar disk dominates in the total surface brightness of the galaxy.
In this region, we can determine the scaling parameters of the exponential shape of the disk surface 
brightness profile \cite{freeman} without including other structural components of the galaxy. 
Decomposition, i.e. a separation of the contributions of the spheroidal and disk
components, is needed in more central regions. The behaviours of the isophote ellipticities 
in the central regions of NGC 524, NGC 502, and NGC 532 are more complex than is expected for 
the superposition of a spheroidal bulge and a flat disk. In NGC 524, the isophote ellipticity 
increases towards the center (Fig.~\ref{iso},b), NGC 532 has a local maximum of the
ellipticity  at $r=4-5^{\prime \prime}$ (Fig.~\ref{iso},c), and NGC 502 has two local 
ellipticity maxima at radii of $4^{\prime \prime}$ and $15^{\prime \prime}$
(Fig.~\ref{iso},a). The position angle also has a distinct value near local maxima 
of the ellipticity: the major axis of the isophotes turns relative to the line of
nodes of the galactic planes. This behaviour of the isophotes implies the presence 
of inner structures in the central regions of these galaxies, such as compact
minibars, or even tilted circumnuclear disks. The large-scale structure of the 
galaxies was analyzed using the GIDRA software \cite{moiseev}, feeded with
the orientation parameters of the isophotes obtained above. The surface brightness profiles 
were constructed by averaging counts in elliptical rings, with a fixed center at the 
galactic nucleus, and with the major axis and ellipticity corresponding to the 
isophotes at a given radius. The resulting profiles were then iteratively
decomposed into an exponential disk (or disks) and a Sersic bulge, starting from the 
outermost regions. This approach is not suitable for galaxies viewed edge-on,
such as NGC 509 and NGC 516; for them, we made two linear cuts, along the major and minor
axes, instead of averaging the surface brightness in rings. The cut along the minor axis 
was then used to estimate the bulge parameters, and the cut along the major axis was 
decomposed into the (already determined) bulge and second-order Bessel function(s)
corresponding to the line-of-sight integral of the surface brightness of a round exponential 
disk that is optically thin up out to its edge. Table 3 presents the parameters of the 
large-scale structural components derived in such a way.

\begin{table*}
\scriptsize
\caption{Parameters of photometric components} 
\begin{flushleft}
\begin{tabular}{|r|c|ccc|ccc|cccc|}
\hline
Galaxy & Band &
\multicolumn{3}{|c|}{Outer disk} & 
\multicolumn{3}{|c|}{Inner disk} & \multicolumn{4}{|c|}{Bulge} \\
& & $\mu _0$, mag/$\square ^{\prime \prime}$ &
$r_0^{\prime \prime}$ &
$r_0$, kpc &
$\mu _0$, mag/$\square ^{\prime \prime}$ &
$r_0^{\prime \prime}$ &
$r_0$, kpc &
n &
$\mu _0$, mag/$\square ^{\prime \prime}$ &
$r_0^{\prime \prime}$ &
$r_0$, kpc \\
NGC~502 & $B$ & 24.6 & 44.7 & 5.2 & 21.2 & 9.9 & 1.2 & 1.5 & 17.1 & 3.4 & 0.4 \\
NGC~502 & $V$ & $\sim$23 & 39.3 & 4.6 & 20.1 & 10.5 & 1.2 & 1.5 & 15.8 & 3.53 & 0.4 \\
NGC~509 & $B$ & 20.6 & 33 & 3.8 & -- & -- & -- & 1.5 & 17.4 & 3.0 & 0.35 \\
NGC~509 & $V$ & 19.6 & 33 & 3.8 & -- & -- & -- & 1.5 & 16.2 & 2.8 & 0.3 \\
NGC~516 & $B$ & 23.0 & 24 & 2.8 & 19.8 & 11.5 & 1.3 & 1.8 & 17.9 & 3.0 & 0.35 \\
NGC~516 & $V$ & 21.9 & 29 & 3.4 & 18.5 & 13 & 1.5 & 1.9 & 16.6 & 2.84 & 0.3 \\
NGC~518 & $B$ & 22.9 & 34.4 & 4.0 & 21.2 & 16.4 & 2.0 & 2 & 17.1 & 8.2 & 0.95 \\
NGC~518 & $V$ & 21.6 & 35.4 & 4.1 & 19.8 & 16 & 1.9 & 2 & 16.4 & 6.67 & 0.8 \\
NGC~524 & $B$ & 21.5 & 30.9 & 3.6 & 19.5 & 9.0 & 1.05 & 1.2 & 17.9 & 3.1 & 0.4 \\
NGC~524 & $V$ & 20.3 & 30.9 & 3.6 & 18.3 & 8.6 & 1.0 & 1.2 & 16.6 & 2.8 & 0.3 \\
NGC~532 & $B$ & 20.8 & 70.2 & 8.2 & 19.45 & 18.7 & 2.2 & 2 & 17.0 & 6.0 & 0.7 \\
NGC~532 & $V$ & 20.45 & 73.2 & 8.5 & 18.2 & 20.5 & 2.4 & 2 & 15.7 & 4.8 & 0.55 \\
\hline
\end{tabular}
\end{flushleft}
\end{table*}

Below we give comments on the decomposition results for the individual galaxies.

\noindent 
{\bf NGC 502.}\\
The center of this galaxy demonstrates a very interesting structure: 
at first glance it seems that two bars of different scales are embedded into 
each other (see also Fig.~\ref{iso},a).
We were able to trace the outer disk of the galaxy from
the radius of $70^{\prime \prime}$: it is a low surface-brightness disk. 
A residual surface-brightness profile obtained after subtracting 
the first model disk from the original image has two ``holes'',
typical for a Freeman Type II profile. We fitted this residual
profile using a second model disk in the interval from 
$23^{\prime \prime}$ to $60^{\prime \prime}$, assuming that there 
is a brightness excess, a ring, between $R=33^{\prime \prime}$ and
$R=45^{\prime \prime}$. After subtracting the second model disk, the residual profile
has a noticeable ``hump'' between $10^{\prime \prime}$ and $24^{\prime \prime}$. 
Basing on the assumption that the galaxy contains two exponential disks and 
a Sersic bulge, we fitted this profile with a somewhat flattened ($b/a = 0.91$)
Sersic bulge with $n=1.5$. The residuals obtained by subtracting the three-component 
model from the initial image suggest that there is a nuclear bar with
a radius of $6.5^{\prime \prime}$ and a ring located approximately between
$5.9^{\prime \prime}$ and $20^{\prime \prime}$, with some shallow brightness peak
in the middle of this interval. Another possibility is that the central structure 
with a radius of $20^{\prime \prime}$ may be a lens; but then there is no place for a bulge in this
galaxy. Further analysis of kinematic data is needed to distinguish between these alternatives.\\

\noindent
{\bf NGC 509.}\\
It appears that this galaxy is viewed edge-on, so that the GIDRA software cannot be used
in this case. Unfortunately, a linear photometric cut can provide less information than 
a brightness profile averaged in rings (ellipses), due to the higher noise level, 
hindering ``extension'' of the profile to large distances from the galactic center. 
It is probably for this reason that we have distinguished only one
exponential disk in the outer regions of this galaxy. The central regions of the profile, 
out to $9^{\prime \prime}$ from the nucleus, were fitted by a Sersic bulge with $n=1.5$.
The brightness profile has a prominent ``hump'', or possibly a flat brightness profile, 
at radii from $9^{\prime \prime}$ to $44^{\prime \prime}$, as is typical of the lenses 
of S0 galaxies. A turn of the isophote major axis by $8^{\circ}$ is visible in
the radial dependency of the position angle at $R=7^{\prime \prime}-30^{\prime \prime}$, 
and the isophote ellipticity shows a maximum (0.7) within this radius range; further, 
out to the edge of the outer disk, the isophote ellipticity drops to 0.4. This region 
($R=7^{\prime \prime}-30^{\prime \prime}$) has a uniformly red colour.
We can treat this inner component as a lens. Or it may be that NGC 509 is not viewed 
quite edge-on, and than the turn of the isophotes indicates that the lens is non-axisymmetric. 
The galaxy apparently possessed structures such as a bar and ring, which have
dissolved somewhat and spread over azimuth by the current moment.\\

\noindent
{\bf NGC 516.}\\
This galaxy is viewed edge-on. Its surface brightness profile demonstrates a bend in the
central region, as is typical for a Freeman Type II profile; this may indicate the presence 
of a bar in the galaxy. The maximum isophote ellipticity (0.7) is observed $20^{\prime \prime}$ 
from the center, supporting the hypothesis that a bar is present. The isophote ellipticity
drops to 0.55 towards the galaxy's edge. However, a thin, blue, edge-on disk ``embedded'' 
into a thicker reddish component is visible in the colour map (Fig.~\ref{bvmaps}),
within about $25^{\prime \prime}$ of the center. It is possible that the
brightness excess in this radius interval is due to the embedded young disk, 
and not to a bar, which would not look uniformly blue along the whole of its length.

We have tried to model the disk component of the galaxy basing on the approach described 
in \cite{erwin08,erwin11}. Note that the NGC 516 profile should be classified as a Type II
profile. As a result, we have extracted outer and inner disks, as well as a bulge that 
dominates the surface brightness within $5^{\prime \prime}$ of the center.\\

\noindent
{\bf NGC 518.}\\
According to the HYPERLEDA data, the galaxy's inclination is $82^{\circ}$. This
is confirmed by the projection of the dust lane against the central region, which 
is clearly visible in the images and colour map. Our isophotal analysis shows 
a constant isophote ellipticity (0.65) starting from $R=35^{\prime \prime}$ and
toward the very edge of the disk. To obtain the surface-brightness profile, we applied 
a photometric cut along the major axis of the isophotes. A ``hump'' is visible
in the profile between the radii of $10^{\prime \prime}$ and $37^{\prime \prime}$ 
which is most likely the result of the intersection of our linear cut with the 
spiral arms. We were able to fit the profile using two exponential disks and a Sersic bulge
with $n=2$.\\

\begin{figure}
\resizebox{\hsize}{!}{\includegraphics{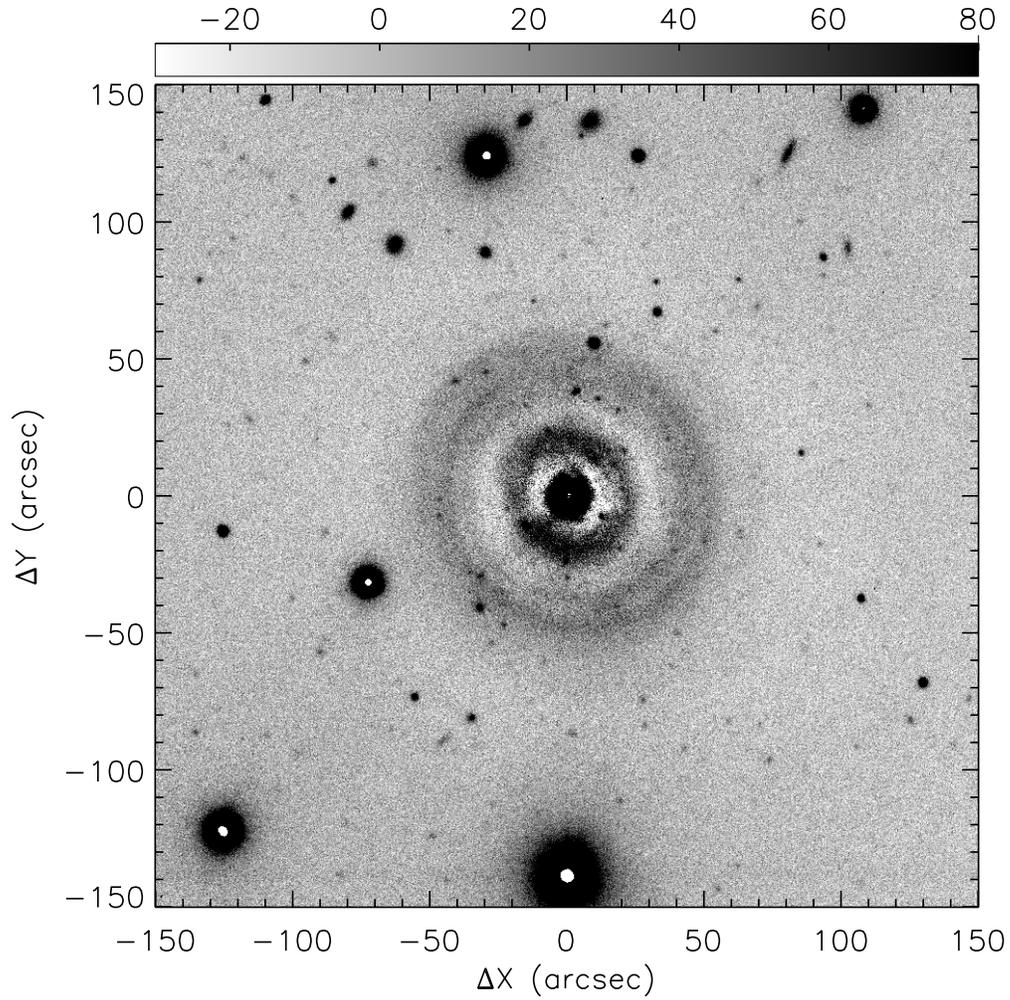}}
\caption{Residual $V$-band image for NGC 524 after subtraction of the 2D three-component model  
(two exponential disks and a Sersic bulge) from the initial image, showing rings at various 
radii in the surface-density distribution of the (old) stellar population.}
\label{n524resi}
\end{figure}

\noindent
{\bf NGC 524.}\\
The outer disk of the central group galaxy can be fitted with a model exponential disk
starting from $70^{\prime \prime}$.  The residual profile between $11^{\prime \prime}$ and 
$35^{\prime \prime}$ should be fitted with an inner exponential disk. Multiple narrow rings
extending to $R\approx 61^{\prime \prime}$ are clearly visible in the residual image. 
We continued to fit the residual image in the radial range of $R=0-11^{\prime \prime}$ with
an almost round Sersic bulge having a quasi-exponential  profile ($n=1.2$).
Subtracting the total three-component model yields residuals containing a
set of weak rings between $34^{\prime \prime}$ and $61^{\prime \prime}$, 
a bright ring at $R=13^{\prime \prime}-25^{\prime \prime}$, and, possibly, an
inclined nuclear disk (Fig.~\ref{n524resi}).

Laurikainen et al. \cite{eija09} recently performed a two-dimensional decomposition of 
a $K_S$-band image of NGC 524 which was less deep than the image analyzed by us here. 
They traced an outer exponential disk to a radius of $80^{\prime \prime}$ and noted 
the presence of two exponential segments with different scales in the residual profile 
within $R=30^{\prime \prime}$. However, their model for the inner region consisted
not of exponential disks, but of a Sersic bulge with n = 2.7 and two Ferrer's lenses 
with flat brightness profiles, every the latter compensating the steep slope of the
bulge brightness profile within the radius ranges of 
$R=0^{\prime \prime}-10^{\prime \prime}$ and $R=10^{\prime \prime}-30^{\prime \prime}$.
It is obvious that, mathematically, decomposition of the two-dimensional
surface-brightness distribution into components admits more than one solution, 
especially if we do not restrict the number of components and their functional
form. A final diagnostics -- (pseudo-)bulge or thin disk dominates in the very center
of NGC 524 -- can be made only by using kinematical data: stellar disks are dynamically 
cooler than spheroids, and their stellar velocity dispersions should be lower than
those in bulges and lenses. We have already carried out a kinematical analysis of the 
stellar component of NGC 524 at various distances from the center, and
the kinematics indicates that the cool disk component dominates within 
$R=10^{\prime \prime}-30^{\prime \prime}$ \cite{we524}.\\

\noindent
{\bf NGC 532.}\\
The fit of the outer parts of this galaxy with an exponential disk can be started from
$R=125^{\prime \prime}$. The residual profile can be characterized as a Freeman's 
Type II profile. We have fitted this residual profile accurately using a model disk so as 
to retain the maximum information. After subtracting two exponential disks, spiral arms 
are clearly visible in the residuals. Further modelling adds a small, strongly flattened
bulge ($b/a = 0.68$) with the Sersic parameter $n=2$. It is difficult to extract details
of the circumnuclear structure due to the high inclination and dust content of the
galaxy.\\

\section{DWARF GALAXIES OF THE GROUP}

As a supplement to the surface photometry of the largest member galaxies 
of the group, we also performed two-colour aperture photometry of smaller 
galaxies within the $6^{\prime}$-fields around the main targets. We carried out
photometry of all objects whose sizes exceeded the estimated full width at half 
maximum of the stellar images. After completing a list of extended objects, the
$B–V$ colour map was made for every field (Fig.~\ref{bvmaps}) and $B-V$ colours
were measured for every list target within a $3^{\prime \prime} - 4^{\prime \prime}$ 
aperture around each. Since most of these galaxies are dwarfs (or compact remote objects), 
their $3^{\prime \prime} - 4^{\prime \prime}$--aperture colours can be considered as the 
integral colours of the galaxies. The accuracy of the measured colours was estimated 
from the pixel-to-pixel scatters within the apertures used for the measurements 
(over $11 \times 11$ or $9 \times 9$ pixels), which ranges from $0.03^m$ for the 
brightest dwarfs to $0.2^m$ for the faintest. The $B–V$ colour index was corrected 
for the extinction in our Galaxy in accordance with \cite{schlegel} recommendations; 
$E(B-V) = 0.08$ was derived from the NED database. In general, 183 galaxies were 
measured in six fields.

Data of the SDSS-III survey covering a section of the sky containing 
the NGC 524 group became available over the Internet at the beginning
of 2011, in particular, integrated $ugriz$ photometry
(so-called ``Petrosian'' magnitudes) became available at the site 
http://skyserver.sdss3.org/dr-8/en/tools/explore/. 
Although the SDSS-III photometry is less deep than our own photometry 
(as we have assured by comparing our $B-V$ colour maps and the SDSS-III $g–r$ 
colour maps for the large galaxies of the group), we have decided to use these 
data in order to separate background galaxies from dwarf members of the group. 
The photometric redshifts for the galaxies having $ugriz$ magnitudes are calculated 
in the SDSS databases. The calculation of photometric redshifts is based on a 
so-called ``learning sample'' -- a list of galaxies for which both photometric 
and spectroscopic redshifts are available. Unfortunately, the learning sample contains 
no dwarf spheroidal or dwarf irregular galaxies, for which very few, if any, spectra 
have been obtained, even for the dwarfs in the nearby Universe, that severely hampers
attempts to determine photometric redshifts for the dwarf members of the NGC 524 group.

Basing on the typical accuracy of photometric redshifts in the SDSS survey, 0.035,
as given in \cite{sdssphotred}, we rejected galaxies with the measured SDSS photometric
redshifts having accuracies better than 0.07 (2$\sigma$) from our list of candidate 
dwarf galaxies of the NGC 524 group. We then verified ``by eye'' the morphology of
the remaining candidates and required that the corrected $B-V$ colour should be bluer 
than 1.3. As a result, 97 objects have remained in the list of dwarf galaxies
of the NGC 524 group, over the six fields analyzed, covering approximate radii of 
30 kpc around the large galaxies of the group. For most of these, we converted
the integrated SDSS $g$ magnitudes to absolute $B$ magnitudes using the relationship 
between the $ugriz$ and $UBVRI$ photometric systems given by \cite{photsyst}; for
the galaxies without SDSS photometry, we measured the $B$ magnitudes using our own data, 
taking fluxes in a series of rising apertures until the background level was reached 
beyond the outermost aperture. We then corrected the $B$ magnitudes for the extinction in our
Galaxy according to the recommendations of NED and converted these into the absolute magnitudes 
using the distance modulus to the group, 31.9 \cite{tonry}. Final results for the entire 
list of the measured dwarf galaxies are given in Table 4.\\

\centerline{Table~4: Coordinates, absolute magnitudes, and the colours for the dwarf
galaxies in the NGC 524 group}
{\scriptsize
\begin{longtable}{|r|p{3.5cm}|l|r|r|}
\hline
No. within the field & $ \alpha_{2000.0}$ & $\delta_{2000.0}$ & $M_B$ &
 $(B-V)_0$ \endhead

\hline
\multicolumn{5}{|l|}{\bf The field of NGC 502}\\
 1 &  01h 23m 04.3s & $+09^{\circ}\,\,\,\, 03^{\prime}\,\,\,\,
 35^{\prime \prime}$  &  --10.92 &  1.26 \\
 2 &  01\quad 22\quad  58.4  &  +09\quad 04\quad 39 & --9.88 & 0.96 \\
 3 &  01\quad 22\quad  48.8  &  +09\quad 04\quad 02 & --10.08 & 0.98 \\
 4 &  01\quad 22\quad  48.0  &  +09\quad 03\quad 49 & --10.50 & 0.69 \\
 5 &  01\quad 22\quad  47.6 &  +09\quad 02\quad 09 & --9.98 & 0.85 \\
 6 &  01\quad 22\quad  52.6 &  +09\quad 00\quad 25 & --8.97 & 1.17 \\
 7 &  01\quad 22\quad  58.2 &  +09\quad 01\quad 39 & --9.85 & 0.68 \\
 8 &  01\quad 23\quad  05.3 &  +09\quad 00\quad 49 & --8.59 & 0.82 \\
 9 &  01\quad 23\quad  05.0 &  +09\quad 00\quad 58 & --9.27 & 1.09 \\
10 &  01\quad 23\quad  04.6 &  +09\quad 00\quad 47 & --9.66 & 0.78 \\
11 &  01\quad 23\quad  00.0 &  +09\quad 00\quad 37 & --9.40 & 1.11 \\
12 &  01\quad 22\quad  53.0 &  +09\quad 03\quad 25 & --10.17 & 0.83 \\
\multicolumn{5}{|l|}{\bf The field of NGC 509}\\
 1 &  01\quad 23\quad  28.4 &   +09\quad 28\quad 01 & --10.55  &  0.63  \\
 2 &  01\quad 23\quad  27.3 &   +09\quad 28\quad 09 &  --8.93  &  0.32  \\
 3 &  01\quad 23\quad  27.9 &   +09\quad 27\quad 21 &  --9.55  &  0.98  \\
 4 &  01\quad 23\quad  23.4 &   +09\quad 27\quad 40 & --11.52  &  0.93  \\
 5 &  01\quad 23\quad  13.1 &   +09\quad 28\quad 01 &  --9.87  &  0.32  \\
 6 &  01\quad 23\quad  16.9 &   +09\quad 27\quad 26 &  --8.81  &  0.36  \\
 7 &  01\quad 23\quad  20.3 &   +09\quad 28\quad 34 &  --9.53  &  0.60  \\
 8 &  01\quad 23\quad  19.8 &   +09\quad 28\quad 29 &  --8.69  &  0.55  \\
 9 &  01\quad 23\quad  20.8 &   +09\quad 27\quad 09 &  --9.03  &  0.34 \\
10 &  01\quad 23\quad  23.7 &   +09\quad 26\quad 59 &  --9.05  &  0.48 \\
11 &  01\quad 23\quad  19.8 &   +09\quad 25\quad 52 & --10.35  &  0.95 \\
12 &  01\quad 23\quad  14.2 &   +09\quad 24\quad 18 &  --9.90  &  0.44 \\
13 &  01\quad 23\quad  24.1 &   +09\quad 25\quad 36 & --10.21  &  0.63 \\
14 &  01\quad 23\quad  24.4 &   +09\quad 24\quad 55 & --10.28  &  0.70 \\
15 &  01\quad 23\quad  28.0 &   +09\quad 24\quad 19 &  --9.25  &  0.58 \\
16 &  01\quad 23\quad  32.3 &   +09\quad 27\quad 18 & --10.32  &  0.66 \\
17 &  01\quad 23\quad  28.6 &   +09\quad 27\quad 42 &  --9.65  &  0.52 \\
18 &  01\quad 23\quad  27.0 &   +09\quad 27\quad 33 &  --9.60  &  1.00 \\
19 &  01\quad 23\quad  21.6 &   +09\quad 26\quad 42 & --10.77  &  0.61 \\
20 &  01\quad 23\quad  19.7 &   +09\quad 27\quad 45 &  --9.58  &  0.38 \\
21 &  01\quad 23\quad  24.6 &   +09\quad 27\quad 01 &  --8.82  &  0.93  \\
22 &  01\quad 23\quad  24.6 &   +09\quad 26\quad 50 & --11.22  &  0.46 \\
\multicolumn{5}{|l|}{\bf The field of NGC 516}\\
 1 & 01\quad 24\quad  11.0 &   +09\quad 35\quad 20 &  --11.29 &   0.86 \\
 2 & 01\quad 24\quad  15.6 &   +09\quad 34\quad 57 &   --9.37 &   1.09 \\
 3 & 01\quad 23\quad  59.6 &   +09\quad 35\quad 20 &   --9.88 &   0.72 \\
 4 & 01\quad 24\quad  05.6 &   +09\quad 31\quad 52 &   --9.71 &   1.00 \\
 5 & 01\quad 24\quad  14.6 &   +09\quad 30\quad 23 &   --9.06 &   1.18 \\
 6 & 01\quad 24\quad  12.7 &   +09\quad 30\quad 24 &   --9.20 &   0.58 \\
 7 & 01\quad 24\quad  06.1 &   +09\quad 30\quad 27 &  --10.67 &   0.94 \\
 8 & 01\quad 24\quad  18.2 &   +09\quad 32\quad 25 &   --9.97 &   0.73 \\
 9 & 01\quad 24\quad  16.0 &   +09\quad 32\quad 19 &   --9.00 &   1.25 \\
10 & 01\quad 24\quad  06.8 &   +09\quad 32\quad 34 &  --11.43 &   1.13 \\
11 & 01\quad 24\quad  06.7 &   +09\quad 32\quad 32 &  --10.97 &   1.10 \\
12 & 01\quad 24\quad  15.6 &   +09\quad 32\quad 51 &   --9.62 &   0.84 \\
13 & 01\quad 24\quad  16.2 &   +09\quad 33\quad 01 &   --9.41 &   0.94 \\
14 & 01\quad 24\quad  16.6 &   +09\quad 33\quad 08 &   --9.39 &   0.76 \\
\multicolumn{5}{|l|}{\bf The field of NGC 518}\\
 1 & 01\quad 24\quad  24.2  &  +09\quad 20\quad 58  &  --9.63  &  0.84 \\
 2 & 01\quad 24\quad  07.6  &  +09\quad 21\quad 06  &  --9.33  &  0.34 \\
 3 & 01\quad 24\quad  22.4  &  +09\quad 18\quad 27  & --11.53  &  0.79 \\
 4 & 01\quad 24\quad  22.6  &  +09\quad 17\quad 17  &  --9.27  &  0.53 \\
 5 & 01\quad 24\quad  27.7  &  +09\quad 18\quad 27  &  --9.71  &  0.43 \\
 6 & 01\quad 24\quad  24.8  &  +09\quad 17\quad 08  &  --8.76  &  0.89 \\
 7 & 01\quad 24\quad  23.8  &  +09\quad 17\quad 30  &  --9.54  &  0.55 \\
 8 & 01\quad 24\quad  10.7  &  +09\quad 22\quad 01  &  --9.81  &  0.74 \\
 9 & 01\quad 24\quad  10.2  &  +09\quad 21\quad 52  &  --8.68  &  0.13 \\
10 & 01\quad 24\quad  08.6  &  +09\quad 20\quad 02  & --12.77  &  0.92 \\
11 & 01\quad 24\quad  17.5  &  +09\quad 21\quad 21  & --11.26  &  0.85 \\
12 & 01\quad 24\quad  20.9  &  +09\quad 22\quad 22  & --10.23  &  1.00 \\
13 & 01\quad 24\quad  26.1  &  +09\quad 17\quad 53  &  --9.07  &  0.89 \\
\multicolumn{5}{|l|}{\bf The field of NGC 524}\\
 1 & 01\quad 24\quad  52.4  &  +09\quad 34\quad 22  &  --10.46 &  1.14 \\
 2 & 01\quad 24\quad  49.7  &  +09\quad 33\quad 49  &  --11.31 &  1.29 \\
 3 & 01\quad 24\quad  54.3  &  +09\quad 34\quad 09  &  --10.23 &  0.06 \\
 4 & 01\quad 24\quad  53.0  &  +09\quad 34\quad 03  &  --12.02 &  0.98 \\
 5 & 01\quad 24\quad  51.9  &  +09\quad 33\quad 52  &  --12.71 &  1.05 \\
 6 & 01\quad 24\quad  42.4  &  +09\quad 34\quad 25  &  --11.57 &  1.16 \\
 7 & 01\quad 24\quad  41.0  &  +09\quad 33\quad 50  &  --11.22 &  0.50 \\
 8 & 01\quad 24\quad  45.6  &  +09\quad 33\quad 38  &  --10.48 &  0.77 \\
 9 & 01\quad 24\quad  53.6  &  +09\quad 31\quad 23  &  --10.38 &  0.78 \\
10 & 01\quad 24\quad  50.7  &  +09\quad 30\quad 51  &  --10.51 &  0.84 \\
11 & 01\quad 24\quad  44.7  &  +09\quad 29\quad 35  &  --10.50 &  0.82 \\
12 & 01\quad 24\quad  44.1  &  +09\quad 29\quad 56  &   --9.94 &  0.10 \\
\multicolumn{5}{|l|}{\bf The field of NGC 532}\\
 1 & 01\quad 25\quad  23.0  &  +09\quad 16\quad 51  &  --9.88  &  1.12 \\
 2 & 01\quad 25\quad  16.8  &  +09\quad 18\quad 28  &  --9.30  &  0.80 \\
 3 & 01\quad 25\quad  14.7  &  +09\quad 18\quad 34  &  --9.31  &  0.66 \\
 4 & 01\quad 25\quad  15.4  &  +09\quad 18\quad 01  &  --8.91  &  0.75 \\
 5 & 01\quad 25\quad  11.8  &  +09\quad 16\quad 52  &  --9.35  &  0.69 \\
 6 & 01\quad 25\quad  08.5  &  +09\quad 15\quad 19  &  --9.39  &  0.25 \\
 7 & 01\quad 25\quad  07.2  &  +09\quad 15\quad 17  &  --9.79  &  1.22 \\
 8 & 01\quad 25\quad  07.6  &  +09\quad 15\quad 02  &  --9.54  &  0.47 \\
 9 & 01\quad 25\quad  21.5  &  +09\quad 13\quad 24  &  --9.70  &  0.98 \\
10 & 01\quad 25\quad  26.7  &  +09\quad 13\quad 22  &  --9.86  &  0.53 \\
11 & 01\quad 25\quad  28.5  &  +09\quad 13\quad 29  &  --9.83  &  0.81 \\
12 & 01\quad 25\quad  22.3  &  +09\quad 13\quad 03  &  --9.27  &  0.09 \\
13 & 01\quad 25\quad  20.2  &  +09\quad 14\quad 37  &  --9.89  &  0.49 \\
14 & 01\quad 25\quad  24.7  &  +09\quad 15\quad 01  &  --9.83  &  0.86 \\
15 & 01\quad 25\quad  17.8  &  +09\quad 14\quad 28  & --10.25  &  0.65 \\
16 & 01\quad 25\quad  20.5  &  +09\quad 13\quad 44  &  --9.59  &  0.49 \\
17 & 01\quad 25\quad  22.5  &  +09\quad 13\quad 43  &  --9.72  &  0.72 \\
18 & 01\quad 25\quad  14.6  &  +09\quad 18\quad 16  &  --9.12  &  1.10 \\
19 & 01\quad 25\quad  09.5  &  +09\quad 18\quad 08  &  --9.16  &  0.08 \\
20 & 01\quad 25\quad  08.9  &  +09\quad 14\quad 04  &  --9.00  &  0.59 \\
21 & 01\quad 25\quad  24.3  &  +09\quad 13\quad 34  &  --9.79  &  0.28  \\
22 & 01\quad 25\quad  23.2  &  +09\quad 13\quad 25  &  --9.15  &  0.65 \\
23 & 01\quad 25\quad  24.2  &  +09\quad 14\quad 10  &  --9.16  &  0.47 \\
24 & 01\quad 25\quad  24.1  &  +09\quad 14\quad 26  &  --9.29  &  0.14 \\
\hline
\end{longtable}
}

\begin{figure*}
\centering
\includegraphics[width=15cm]{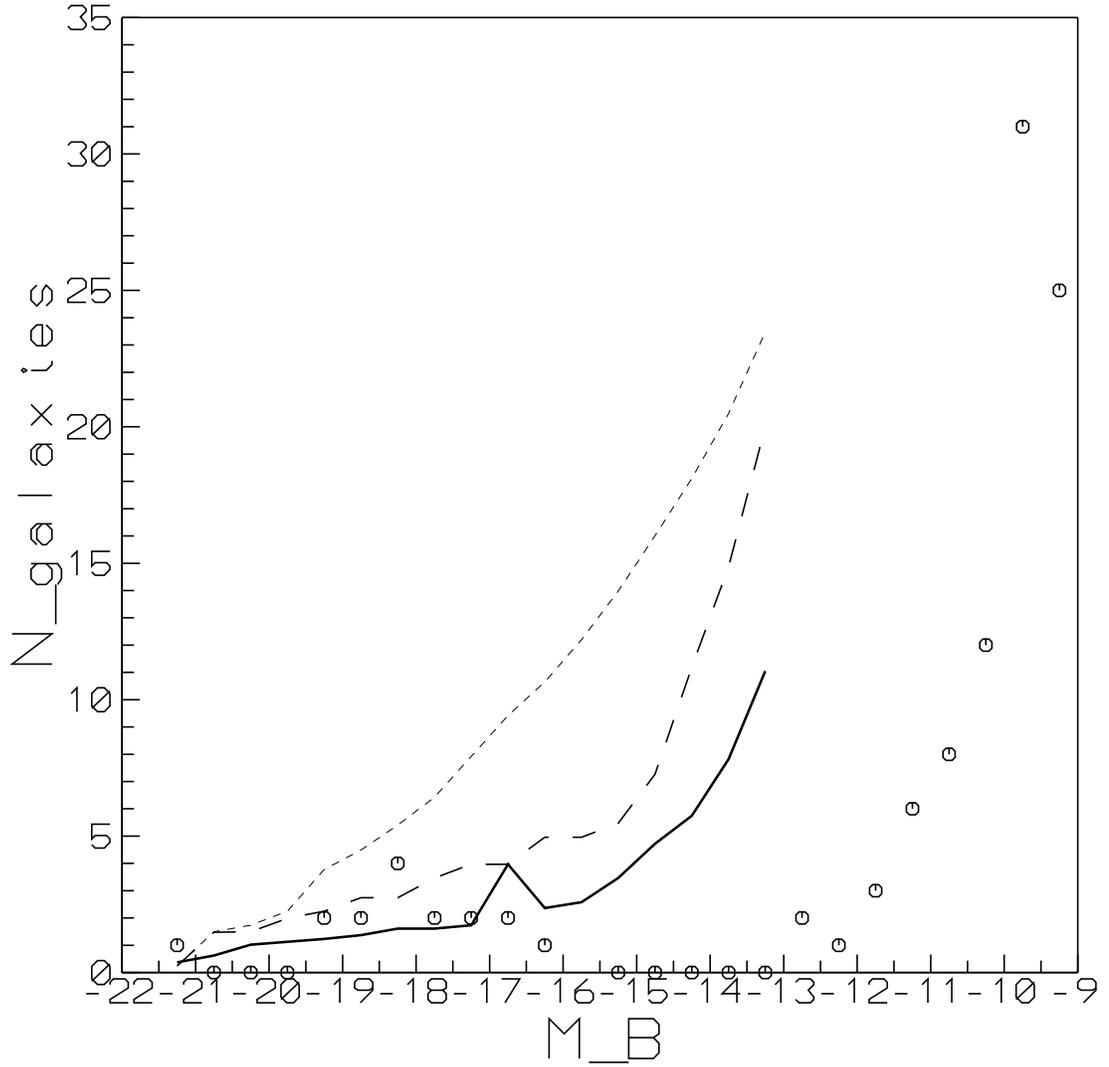}
\caption{Luminosity function of the galaxies of the NGC 524 group in the range of
$M_B = -22$ to $-16$, following the data of [3],
and in the range of $M_B = -13$ to $-9$, following our measurements. 
For comparison, lines show the average luminosity functions of groups of different masses, 
according to [21]: massive groups with high X-ray luminosities (dotted line), 
groups with intermediate masses and low X-ray luminosities (dashed line), and groups without 
X-ray emitting gas (solid line).}
\label{lumfunc}
\end{figure*}

It is astonishing that our lists contain no galaxies brighter than $M_B = -13^m$, 
and such galaxies were also not found in wider neighbourhoods of the studied
fields, which we surveyed using a navigator of the eighth SDSS data release. 
Since the NED database does not detect also any such galaxies during a
search for members of the group over the total area of the group, we conclude 
that there exists no group galaxies with the absolute magnitudes between $M_B=-13^m$
and $-16^m$, i.e., the luminosity function of the group galaxies is clearly bimodal.
Figure~\ref{lumfunc} presents the luminosity function of the identified dwarfs and 
of the 16 large members of the group, for which we take the integrated blue
magnitudes from \cite{vennik}. Formally, we must match the two parts  of the 
luminosity function -- this for large and that for small galaxies, with a 
boundary near $M_B=-16^m$ -- by multiplying the dwarf branch by a factor related 
to the fact that we have not studied the full area of the group. Assuming 
a uniform distribution of dwarf galaxies over the group area, this factor 
should be 35. However, Fig.~\ref{lumfunc} shows that so substantial
correction is not needed, and the required factor does not exceed 10: 
apparently, the dwarfs are concentrated near the large galaxies, being 
satellites of the latter. Figure~\ref{lumfunc} also compares the luminosity 
function of the members of the NGC 524 group with the average luminosity 
functions of various-mass groups calculated in \cite{lumfunc_group}. The slope of 
the dwarf branch is consistent with all the average luminosity functions 
(although it is shifted by several magnitudes toward the area of faint galaxies). 
In the bright part of the luminosity function, the number of galaxies in the 
NGC 524 group matches the luminosity functions of groups with zero or low X-ray 
luminosities (with low or intermediate total masses); however a bimodal luminosity 
function is exclusively a feature of groups without X-ray emitting
gas at all \cite{lumfunc_group}. The ROSAT data indicate an X-ray luminosity
of $\log (L_X \mbox{[erg/s]}) = 41.05$ for the NGC 524 group \cite{xray},
but there is a note that the X-ray emission is tightly concentrated around the 
central galaxy, NGC 524. Now we see that the shape of the galactic luminosity function 
in the NGC 524 group supports the idea that the group itself {\it has no} hot 
intergalactic gas.

\begin{figure*}
\centering
\includegraphics[width=15cm]{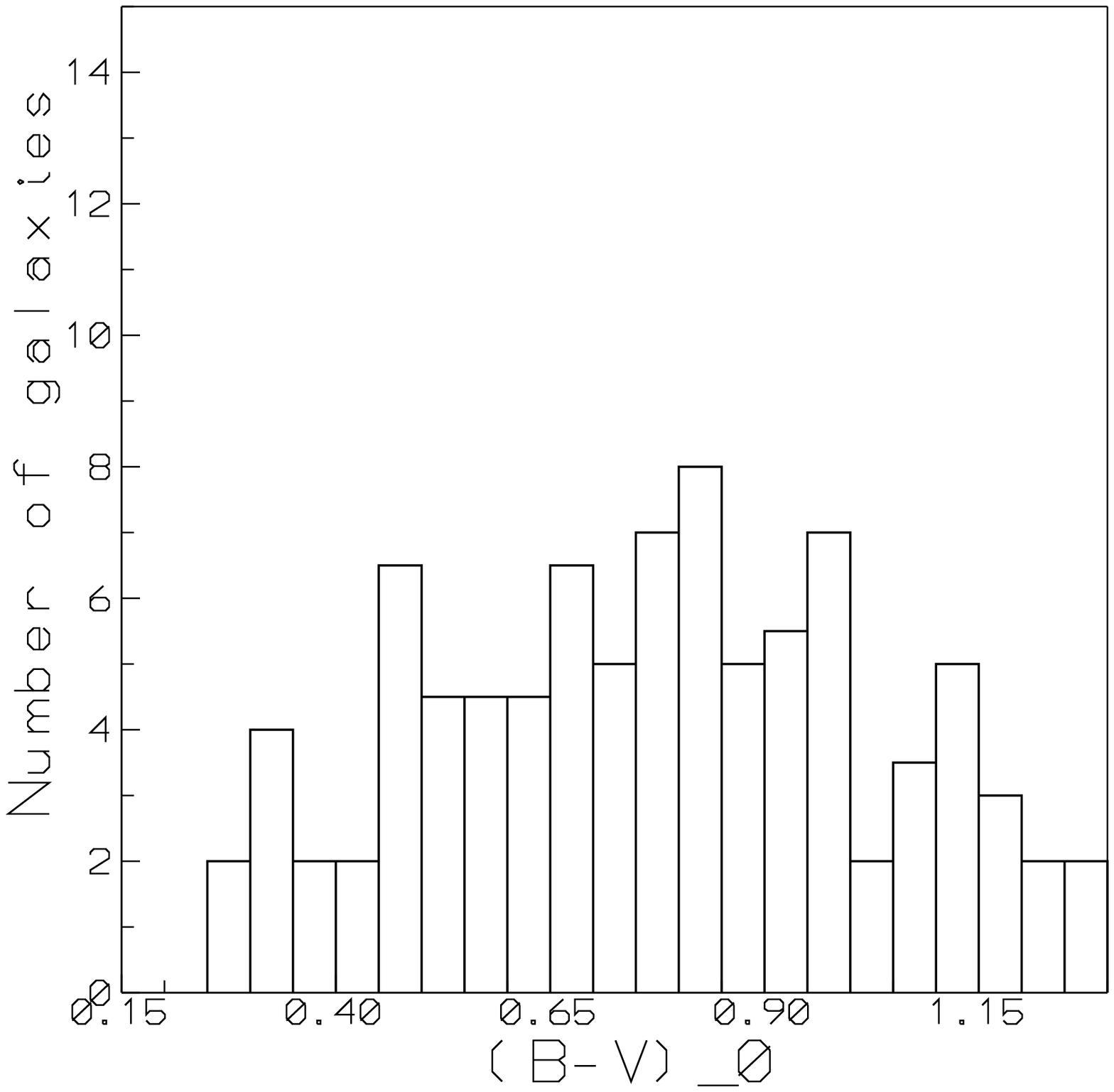}
\caption{Distribution of dwarf galaxies of the NGC 524 group over their $B-V$ 
colours (corrected for the reddening in our Galaxy).}
\label{bvhisto}
\end{figure*}

Figure~\ref{bvhisto} presents the $B-V$ colour distribution for the dwarf galaxies. 
This distribution is nearly flat in the colour interval of $B-V = 0.4-1.0$: we do not
see the usual bimodality, with a ``red sequence'' and ``blue cloud''. Although 
we could suppose, based on the example of our Local Group, that red dwarfs must be
concentrated near large host galaxies and blue dwarfs are mostly located far from 
such galaxies: as a rule, tidal effects from large galaxies must suppress star formation
in their satellites over dynamical timescales. However, in the case of the NGC 524 group, 
a considerable number of blue dwarfs with star formation were found just near the large galaxies.

\begin{figure}
\resizebox{\hsize}{!}{\includegraphics{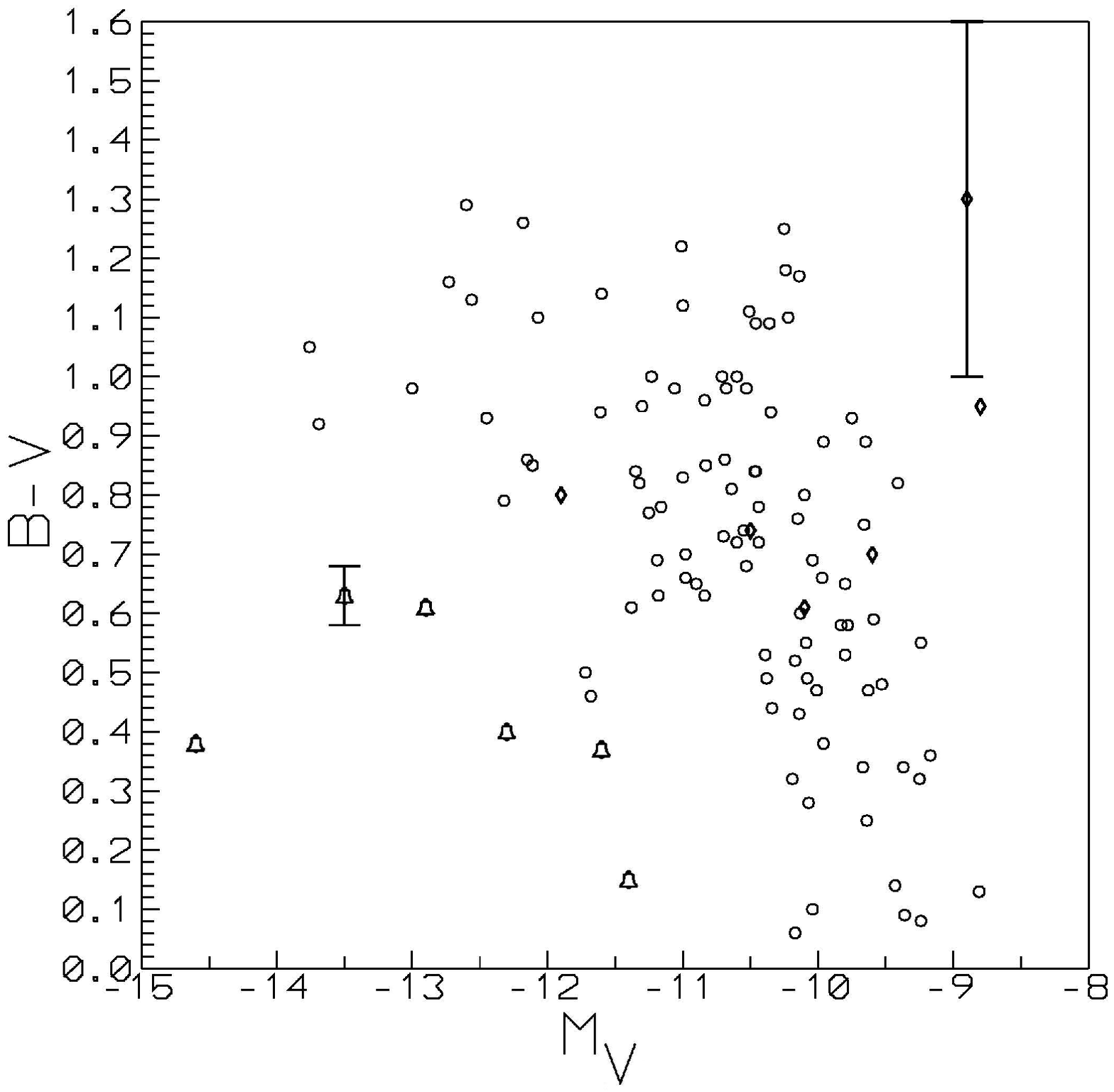}}
\caption{Colour-luminosity diagram for the dwarf galaxies of the NGC 524 group compared 
to dwarf galaxies of the Local Group. The circles show dwarf galaxies of the NGC 524 group, 
the bells -- late-type dwarfs of the Local Group with young stellar populations, and the 
diamonds -- dSph dwarfs of the Local Group with old stellar populations}
\label{bvdistri}
\end{figure}

The colour--luminosity diagram in Fig~\ref{bvdistri} compares the dwarf populations of 
the NGC 524 group and that of the Local Group (the data for the Local Group were taken
from \cite{lgreview}). Dwarfs of both early (dSph) and late (dIrr) types were selected 
in the Local Group, and are plotted in the colour-luminosity diagram together with
the members of the NGC 524 group. The dwarfs of the Local Group and NGC 524 group occupy 
approximately the same colour interval; the reddest dwarf of the Local Group, Ursa Minor, is
a dSph dwarf with $B-V = 1.3 \pm 0.3$, like the reddest dwarfs of the NGC 524 group. However, 
the bluest dwarfs of the Local Group are more luminous than the bluest dwarfs found near the large 
galaxies of the NGC 524 group, on average by approximately $2^m$.

\section{CONCLUSIONS}

We have analyzed the structures of the six largest
disk galaxies of the NGC 524 group using surface
photometry data obtained at the SAO 6-m telescope with
the SCORPIO reducer. Four of the galaxies are classified in
the literature as lenticular galaxies, and another two
are classified as early-type spiral galaxies; two galaxies
are viewed almost face-on, while the remaining
four are inclined at large angles to the line of sight.
As far as we can judge from their images, none of
the galaxies viewed not quite edge-on has a large-scale
bar. According to the behaviour of the radial surface-brightness
distributions, the only galaxy that may contain
a bar is NGC 516, which is viewed strictly edge-on;
but, more probably, it is not a bar but an embedded,
thin inner edge-on disk. In general, bars are
present in 40\%-–70\%\ of disk galaxies; their absence
in all the large galaxies of the NGC 524 group could
indicate some influence from their environment.
However, we detected both inner and relative large outer
rings in almost all the large disk galaxies in the
NGC 524 group. In the face-on galaxies NGC 502
and NGC 524, rings of different radii fill out almost the
entire range of radial distances. The nearby statistics of 
the frequency of bars in disk galaxies indicates that about 
50\%\ of such galaxies have inner rings \cite{innrings}, 
while 20\%\ have nuclear rings (with radii out to 1.5 kpc) 
\cite{nucrings}, with a bar being present in the overwhelming 
majority of galaxies with rings. The disk galaxies of the NGC 524 
group have rings but no bars. It is interesting that no neutral hydrogen
was found in the four lenticular galaxies of the group \cite{sengupta}; 
i.e., there is no gas and so no star formation,
but there are rings (waves) of surface brightness. If
such structural anomalies are to be explained by an
enhanced contribution of secular evolution during the
formation of the large-scale structure of the galaxies
in a dense environment, we must search for special mechanisms
of the secular evolution
that would not switch on star formation (for which there
is no fuel in this case). Here we must also explain why
the Sersic exponents of the bulges of all the group
members are lower than two: bulges built via the
secular evolution of disks with bars and gas may have exponential
bulges, while ``minor merging'' (the capture of satellites) 
usually increases the Sersic exponents in bulges to 3--4 \cite{aguerri}.
The presence of a large number of the blue dwarf galaxies undergoing 
current star formation within some 20--30 kiloparsec from the large galaxies 
of the group is interesting. Dynamical evolution over scales of a
few billion years should cause these dwarfs to ``fall'' into their parent 
galaxies, or at least deprive them of gas and ongoing star formation, 
as in the Local Group. Thus, the systems of blue dwarfs around large
red galaxies should either be systems recently assembled, or should be 
in stable orbits that prevent their falling into the centers of parent galaxies; 
otherwise, the continuous accretion of gas-rich dwarfs onto the disks 
of lenticular galaxies would provide the large galaxies with a fuel 
for ongoing star formation. Our spectral studies have shown that the 
outer disks of NGC 502 and NGC 524 have very old stellar populations \cite{we12},
so this accretion does not occur for some reason.

\bigskip

The data used in this study were obtained at the
6-m telescope of the Special Astrophysical Observatory
of the Russian Academy of Sciences, operated
with financial support from the Ministry of Education
and Science of the Russian Federation (registration
number 01-43). We thank A.V. Moiseev for supporting
these observations. The data analysis made use of
the Lyon-Meudon Extragalactic Database (LEDA),
maintained by the LEDA team at the Lyon Centre
for Astrophysics Research (CRAL; France), and the
NASA/IPAC Extragalactic Database (NED), operated
by the Jet Propulsion Laboratory of the California
Institute of Technology under contract with
the National Aeronautics and Space Administration
(USA). We also used public data of the SDSS-III
project (http://www.sdss3.org), supported by the Alfred
P. Sloan Foundation, the participating institutions
of the SDSS Collaboration, the National Science
Foundation, and the U.S.Department of Energy
Office of Science. This study was supported by the
Russian Foundation for Basic Research 
(project no. 10--02--00062a).

\end{document}